\documentclass[a4paper,11pt]{article}
\usepackage{pos}
\usepackage{hyperref}
\usepackage{graphicx}

\title{Light Baryon Spectroscopy at BESIII}
\author*[a,b]{Hao Liu}
\author[a,b]{Xiongfei Wang}
\author[c]{Haibo Li}

 \onbehalf{on behalf of  BESIII Collaboration}

\affiliation[a]{School of Physical Science and Technology, Lanzhou University, Lanzhou 730000, People’s Republic of China}

\affiliation[b]{
Lanzhou Center for Theoretical Physics,
Key Laboratory of Theoretical Physics of Gansu Province,
Key Laboratory of Quantum Theory and Applications of MoE,
Gansu Provincial Research Center for Basic Disciplines of Quantum Physics,
Lanzhou University, Lanzhou 730000, People’s Republic of China
}

\affiliation[c]{Institute of High Energy Physics, Chinese Academy of Sciences, No. 19B Yuquan Rd., Shijingshan District, Beijing, the People's Republic of China}
\emailAdd{lhao2023@lzu.edu.cn}
\abstract{
The BESIII collaboration has collected large data samples in the center-of-mass energy range from 1.84 to 4.95 GeV. These data provide a low-background environment and suitable phase space for studying the spectroscopy of light baryons. In this article, we review the achievements of baryon spectroscopy studies by the BESIII collaboration. Most of the results are obtained through the partial wave analysis (PWA) method, with spin and parity well determined.
}

\FullConference{
The 21st International Conference on Hadron Spectroscopy and Structure (HADRON2025)
}

\begin{document}
\maketitle

\section{Introduction}
The quark model~\cite{Gell-Mann:1964ewy} requires that a baryon consists of three quarks. Based on ${\rm SU}(3)$ symmetry with three light quarks $u$, $d$ and $s$, we could define the baryonic octet and dectet. 
At present, all octet and decuplet baryons have been confirmed well, but for the much higher excited states, we are still lacking knowledge. Experimentally in the Particle Data Group (PDG)~\cite{ParticleDataGroup:2024cfk}, numerous excited states of baryons have been recorded with different evidence of existences, with the improvement of detection technique and the expanded energy limit, more states will be discovered and updated accurately. Theoretically, several studies have explored the behaviors by Lattice QCD~\cite{Edwards:2011jj}, Relativistic quark model~\cite{Thiel:2022xtb} and Regge trajectory~\cite{Menapara:2021dzi}, furthermore, for these calculated results which are not confirmed well in experiment, many of the unique baryon models are raised up, such as quark-diquark model, Y/$\Delta$-type model and pentaquark state model, etc. Due to the quark confinement, such models could not be determined directly, so that it is better to study via hadronic interaction and decays. Partial wave analysis (PWA) is a method for the hadron spectroscopy study via amplitude analysis, which could determine the properties much quantitatively including mass, width, spin, and parity, etc. To date, many experiential collaborations have accepted PWA as a general method to study hadron spectroscopy, especially baryon. Compared with the other collaborations, the BESIII collaboration has its unique advantages for baryon spectroscopy study. Firstly, the BESIII detector is located on a symmetric $e^+ e^-$ collider BEPCII, which could bring a low background and high signal via $e^+ e^-$ annihilating; secondly, the energy range for BESIII is very wide so that it could cover most of light baryon production with an appropriate phase space, which could contribute a greater cross section of baryon production; thirdly, BESIII collaboration has collected substantial data samples, such as about 10 billion for $J/\psi$ events, 2.7 billion for $\psi(2S)$ events, 20 ${\rm fb}^{-1}$ $\psi(3770)$ sample and other $XYZ$ data sample with scanned larger than 500 ${\rm pb}^{-1}$ per energy point at 10-20 MeV interval, which is a huge statistical sample in the world. With the advantages in the BESIII detector, the collaboration has reported substantial results for light baryon spectroscopy, and in this article, we will introduce the highlights of baryon spectroscopy on BESIII in the recent years.

\section{The BESIII Detector}
The BESIII detector~\cite{Ablikim:2009aa} records symmetric $e^+e^-$ collisions provided by the BEPCII storage ring~\cite{Yu:IPAC2016-TUYA01} in the center-of-mass energy range from 1.84 to 4.95 GeV, with a peak luminosity of $1.1 \times 10^{33}\;\text{cm}^{-2}\text{s}^{-1}$ achieved at $\sqrt{s} = 3.773\;\text{GeV}$. BESIII has collected large data samples in this energy region~\cite{Ablikim:2019hff,EcmsMea,EventFilter}. The cylindrical core of the BESIII detector covers 93\% of the full solid angle and consists of a helium-based multilayer drift chamber~(MDC), a time-of-flight system~(TOF), and a CsI(Tl) electromagnetic calorimeter~(EMC), which are all enclosed in a superconducting solenoidal magnet providing a 1.0~T magnetic field (0.9~T in 2012). The solenoid is supported by an octagonal flux-return yoke with resistive plate counter muon identification modules interleaved with steel. The acceptance of charged particles and photons is 93\% over the $4\pi$ solid angle. The charged-particle momentum resolution at $1~{\rm GeV}/c$ is $0.5\%$, and the ${\rm d}E/{\rm d}x$ resolution is $6\%$ for electrons from Bhabha scattering. The EMC measures photon energies with a resolution of $2.5\%$ ($5\%$) at $1$~GeV in the barrel (end cap) region. The time resolution in the plastic scintillator TOF barrel region is 68~ps, while that in the end cap region was 110~ps.  The TOF end cap system was upgraded in 2015 using multi-gap resistive plate chamber technology, providing a time resolution of 60~ps, which benefits the data quality~\cite{etof1,etof2,etof3}.

\section{Highlights on BESIII}
\subsection{$N$ Baryon}
\begin{figure}[!htp]
    \centering
    \includegraphics[scale=0.368]{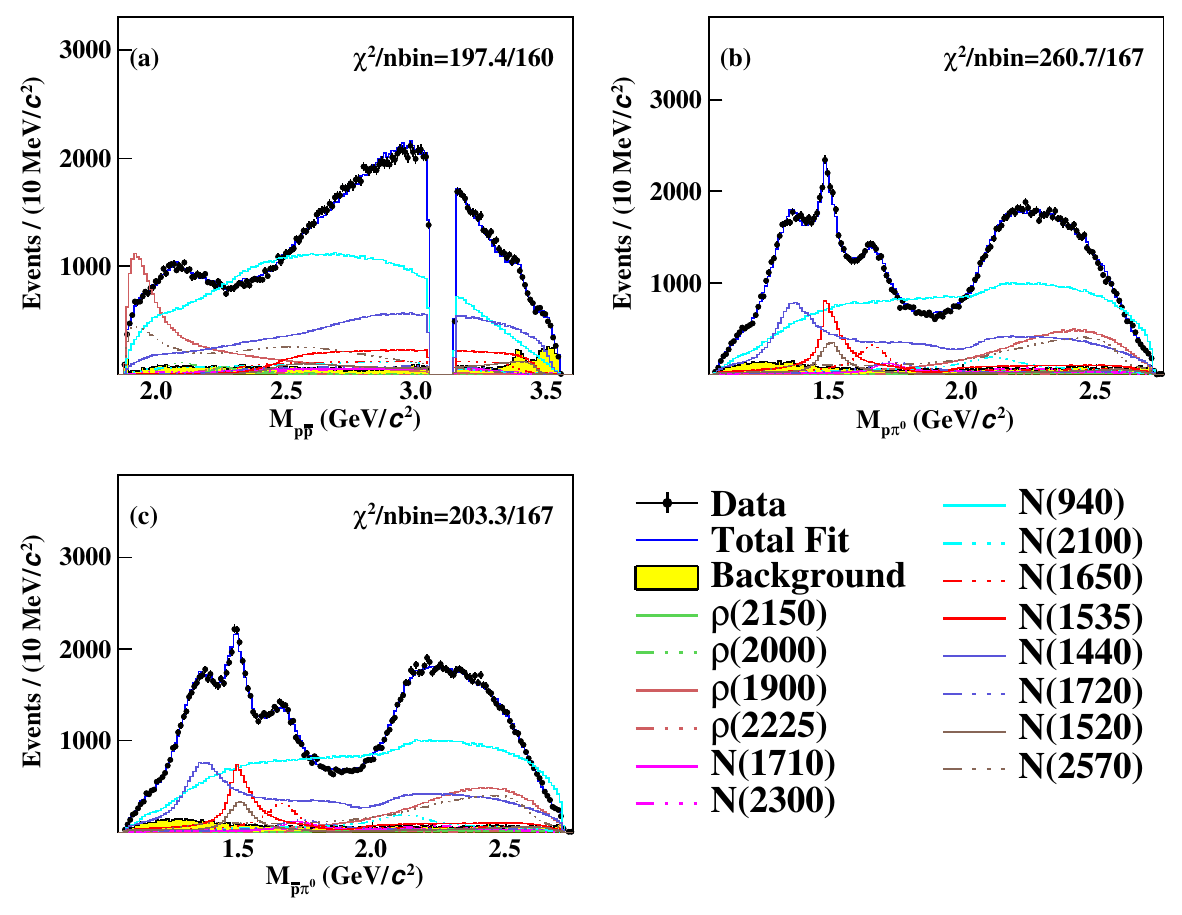}
    \includegraphics[scale=0.368]{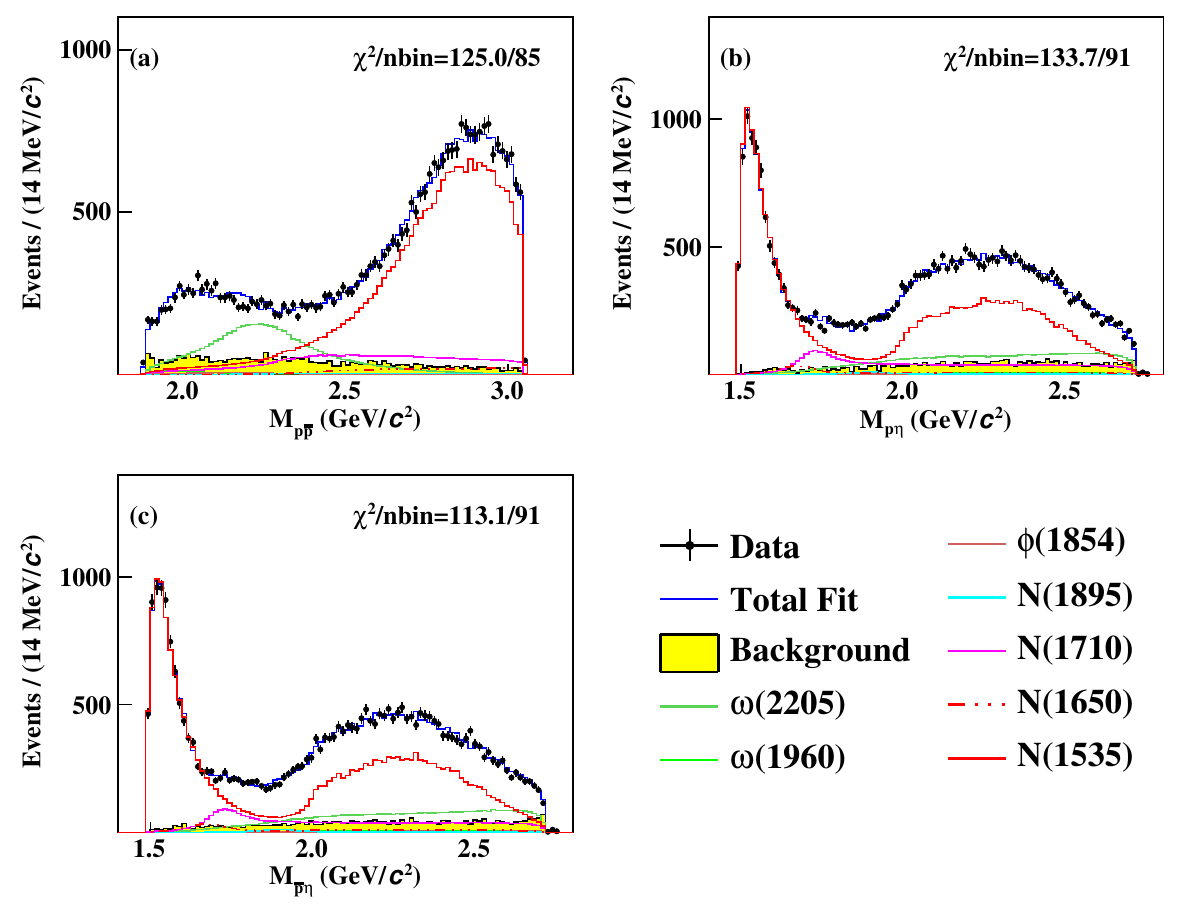}
    \caption{PWA fit curves for $\psi(3686) \to p \bar{p} \pi^0$ (left) and $p \bar{p} \eta$ (right)~\cite{BESIII:2024vqu}.}
    \label{fig:pwaPPX}
\end{figure}

BESIII recently reported on the $N^*$ and other excited mesons via $\psi(3686) \to p \bar{p} \pi^0$ and $p \bar{p} \eta$ using a data sample of 2.7 B $\psi(3686)$~\cite{BESIII:2024vqu}, a PWA method is performed in this study, and all resonances are formulated by the Breit-Wigner (BW) function with an energy-dependent width $\Gamma$~\cite{Hunt:2018wqz}. In this analysis, except for several $N^*$ resonances, the other states of the meson over $1.8$ GeV are determined at the same time. There are no additional states discovered compared to PDG~\cite{ParticleDataGroup:2024cfk}. The PWA fit curves and resonances are summarized in Fig.~\ref{fig:pwaPPX} and Table~\ref{tab:baselineN}, respectively.
\begin{table}[!htp]
	\centering
	\caption{The resonance parameters, quantum numbers in the baseline solutions $\psi(3686)\to p\bar{p}\pi^0$ and $\psi(3686)\to p\bar{p}\eta$~\cite{BESIII:2024vqu}.}
\scalebox{0.67}{
	\begin{tabular}{lccc|lccc}
		\hline
		\hline
		Resonance state & Mass (MeV/$c^2$) & Width (MeV) & $I^G(J^{PC})$ &Resonance state & Mass (MeV/$c^2$) & Width (MeV) & $I^G(J^{PC})$ \\
  \hline
		$N(940)$  &$940$  &$0$ & $1/2(1/2^+)$ &$\rho(1900)$  &  $1880\pm30$  & $130\pm30$ & $1^+(1^{--})$ \\
		$N(1440)$&  $1406\pm3$& $314\pm9$  & $1/2(1/2^+)$ &$\rho(2000)$  &  $2078\pm6$  & $149\pm21$ & $1^+(1^{--})$ \\
		$N(1520)$ &  $1512\pm2$& $121\pm3$  & $1/2(3/2^-)$ &$\rho(2150)$  &  $2254\pm22$  & $109\pm76$ & $1^+(1^{--})$ \\
		$N(1535)$ & $1525\pm2$ & $147\pm5$  & $1/2(1/2^-)$ &$\rho(2225)$  &  $2225\pm35$  & $335\pm100$ & $1^+(2^{--})$ \\
		$N(1650)$ & $1666\pm3$  & $133\pm7$  & $1/2(1/2^-)$ &$\phi_3(1850)$  &  $1854\pm7$  & $87\pm28$ & $0^-(3^{--})$ \\
		$N(1710)$ & $1710\pm30$ & $140\pm60$  & $1/2(1/2^+)$ &$\omega(1960)$  &  $1960\pm25$  & $195\pm60$ & $0^-(1^{--})$ \\
		$N(1720)$ & $1720\pm30$ & $250\pm100$  & $1/2(3/2^+)$ &$\omega(2205)$  &  $2205\pm30$  & $350\pm90$ & $0^-(1^{--})$ \\
		$N(1895)$ & $1895\pm25$  & $120\pm80$  & $1/2(1/2^-)$ \\
		$N(2100)$ & $2100\pm50$  & $260\pm60$  & $1/2(1/2^+)$ \\
		$N(2300)$ & $2300\pm116$  & $340\pm114$  & $1/2(1/2^+)$\\
		$N(2570)$ & $2570\pm39$  & $250\pm70$  & $1/2(5/2^-)$\\
		\hline
		\hline
	\end{tabular}
 }
	\label{tab:baselineN}
\end{table}
\subsection{$\Lambda$ and $\Sigma$ Baryon}
For $\Lambda$ baryon, there are several resonances detected via PWA method, such as $\Lambda(1670)$ in $\psi(3686) \to \Lambda \bar{\Lambda} \eta$~\cite{BESIII:2022cxi}, $\Lambda(2000)$ in $\psi(3686) \to \Lambda \bar{\Lambda} \omega$~\cite{BESIII:2022fhe} and $\Lambda(2325)$ in $J/\psi \to \Lambda \bar{\Sigma}^0 \pi^0$~\cite{BESIII:2024jgy}. The significances and baseline solutions are listed in the following Table~\ref{SummaryLambda}, where $\Lambda(2000)$ is inconsistent with the PDG result, while the significance is only $3.0\sigma$ for the leak of statistics; Furthermore, in Ref.~\cite{BESIII:2024jgy}, the $\Lambda(1405)$ state is parameterized as the Flatt\'e-like~\cite{Chung:1995dx} and chiral dynamics~\cite{Jido:2003cb} models independently, while due to limited statistics, two models could not be determined which one is much optimal so that both parameterized results have been reported in this analysis. For the study of $e^+ e^- \to p K^- \bar{\Lambda}$ at 4.178 GeV~\cite{BESIII:2023kgz}, no additional resonances $\Lambda^*$ were discovered except a new $1^+$ resonance named as $K_1(2085)$ with $M=2084_{-3}^{+4} \pm 25$ MeV/$c^2$ and $\Gamma = 58^{+4}_{-3} \pm 25$ MeV in $p\bar{\Lambda}$ spectrum. For $\Sigma$ baryon, the BESIII collaboration has studied in $J/\psi \to \Sigma^{\mp} \bar{\Lambda} \pi^{\pm}$~\cite{BESIII:2023syz} and mentioned $J/\psi \to \Lambda \bar{\Sigma}^0 \pi^0$~\cite{BESIII:2024jgy}, several resonances $\Sigma^*$ are considered in the nominal fit according to PDG~\cite{ParticleDataGroup:2024cfk}, but no new states have been confirmed.
\begin{figure}[!htp]
    \centering
    \includegraphics[scale=0.245]{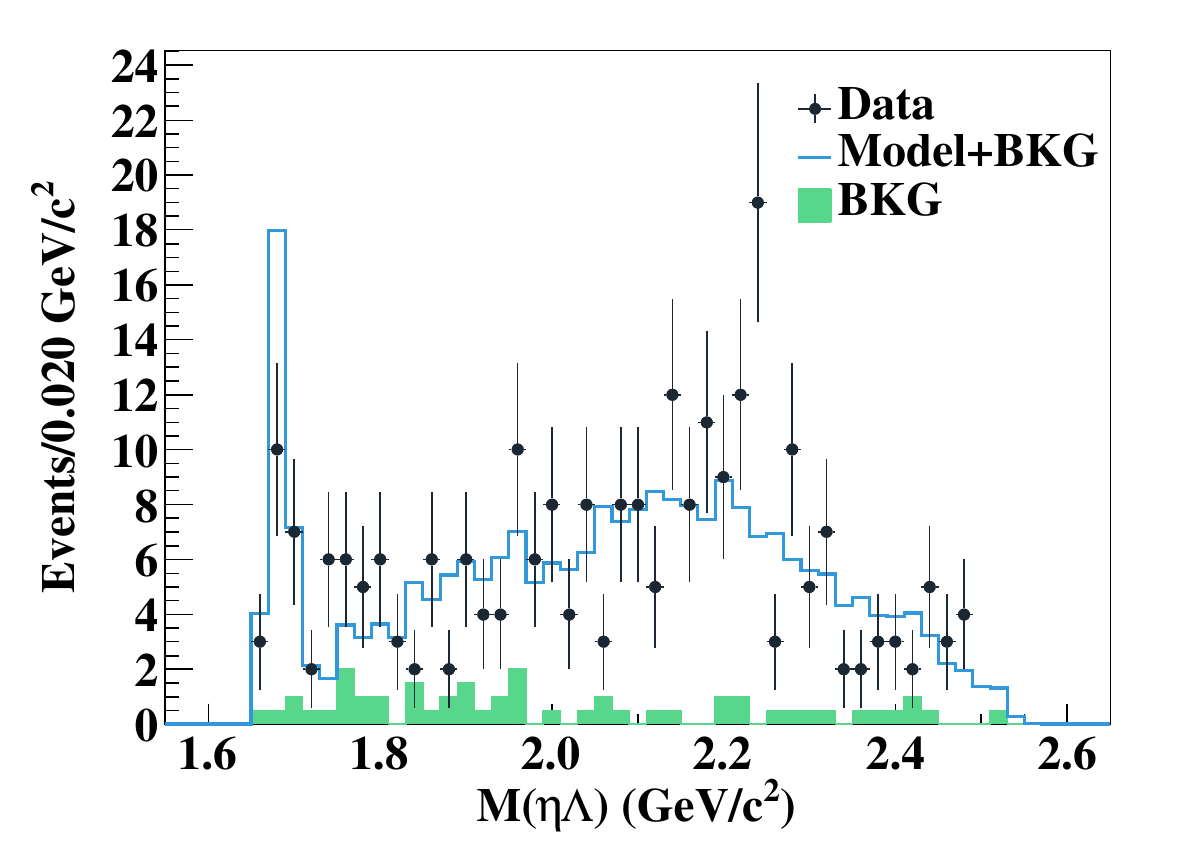}
    \includegraphics[scale=0.245]{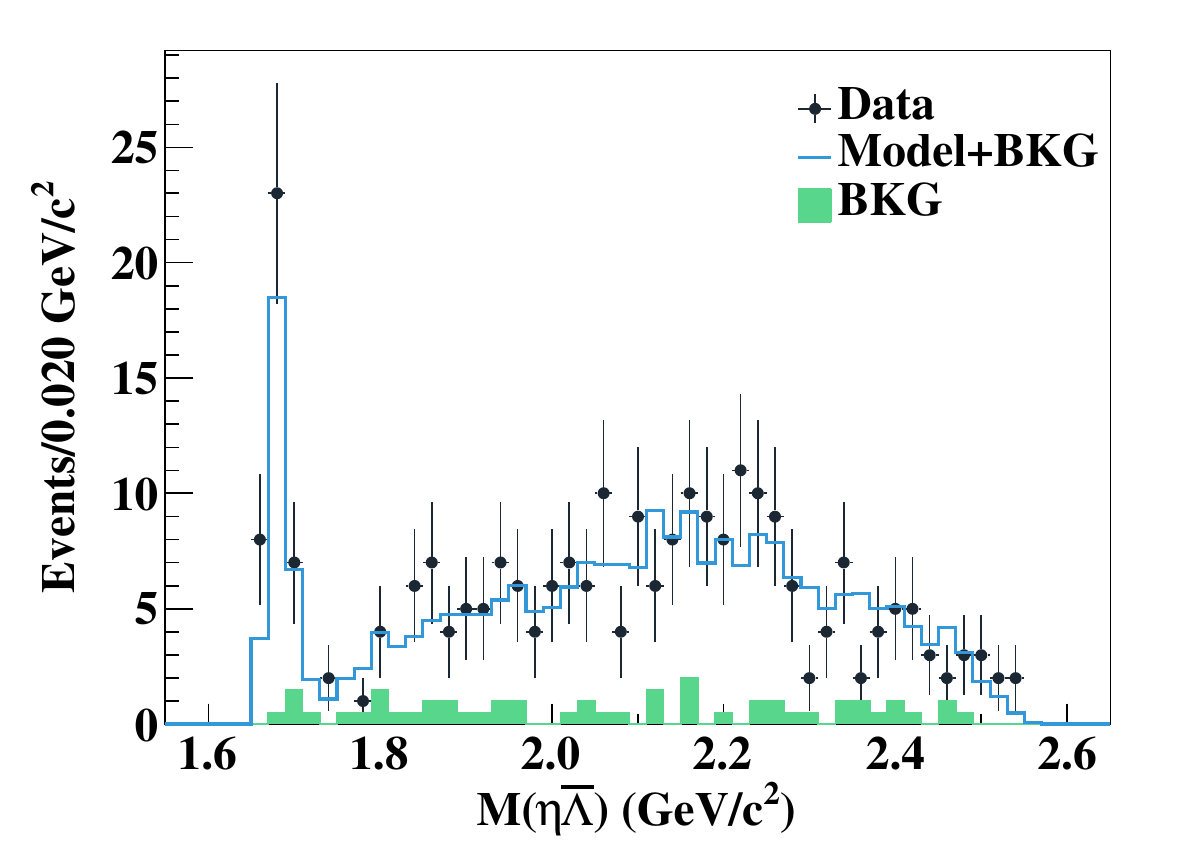}
    \includegraphics[scale=0.245]{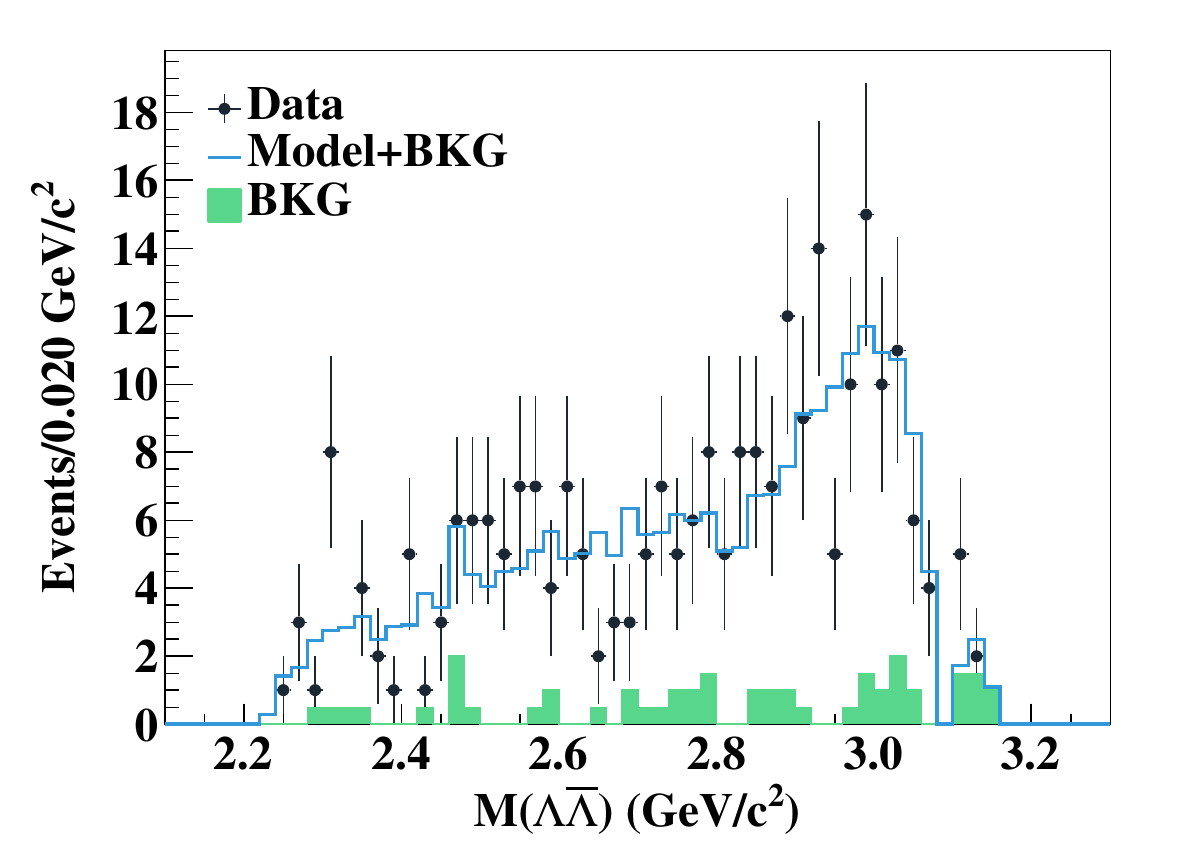}
    \caption{PWA fit curves for $\psi(3686) \to \Lambda \bar{\Lambda} \eta$~\cite{BESIII:2022cxi}.}
    \label{fig:pwaLLE}
\end{figure}
\begin{figure}[!htp]
    \centering
    \includegraphics[scale=0.75]{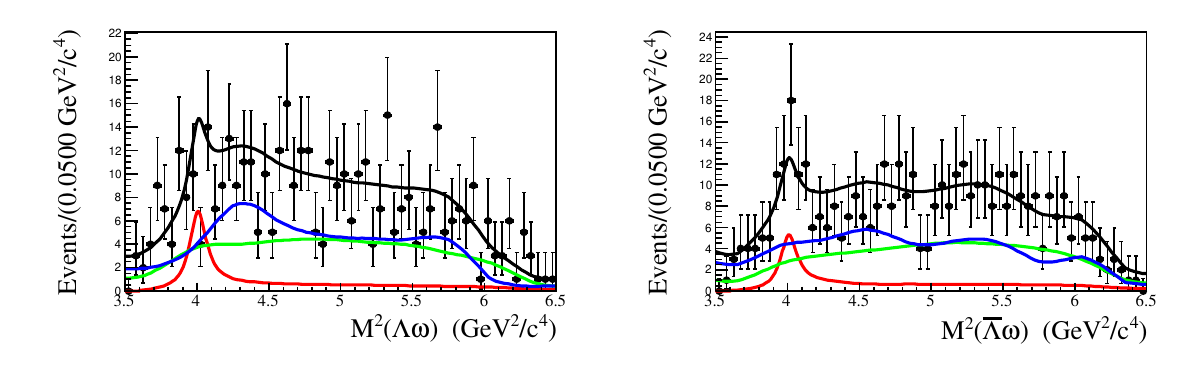}
    \caption{PWA fit curves for $\psi(3686) \to \Lambda \bar{\Lambda} \omega$~\cite{BESIII:2022fhe}.the black solid curves are projections from the fits, the red solid curves show the shape of $\Lambda^*/\bar{\Lambda}^*$ resonance, the blue solid curves show the background described by $\omega$ sidebands and the green solid curves show the shapes from the non resonant decay $\psi(3686) \to \Lambda \bar{\Lambda} \omega$.}
    \label{fig:pwaLLO}
\end{figure}
\begin{figure}[!htp]
    \centering
    \includegraphics[scale=0.3]{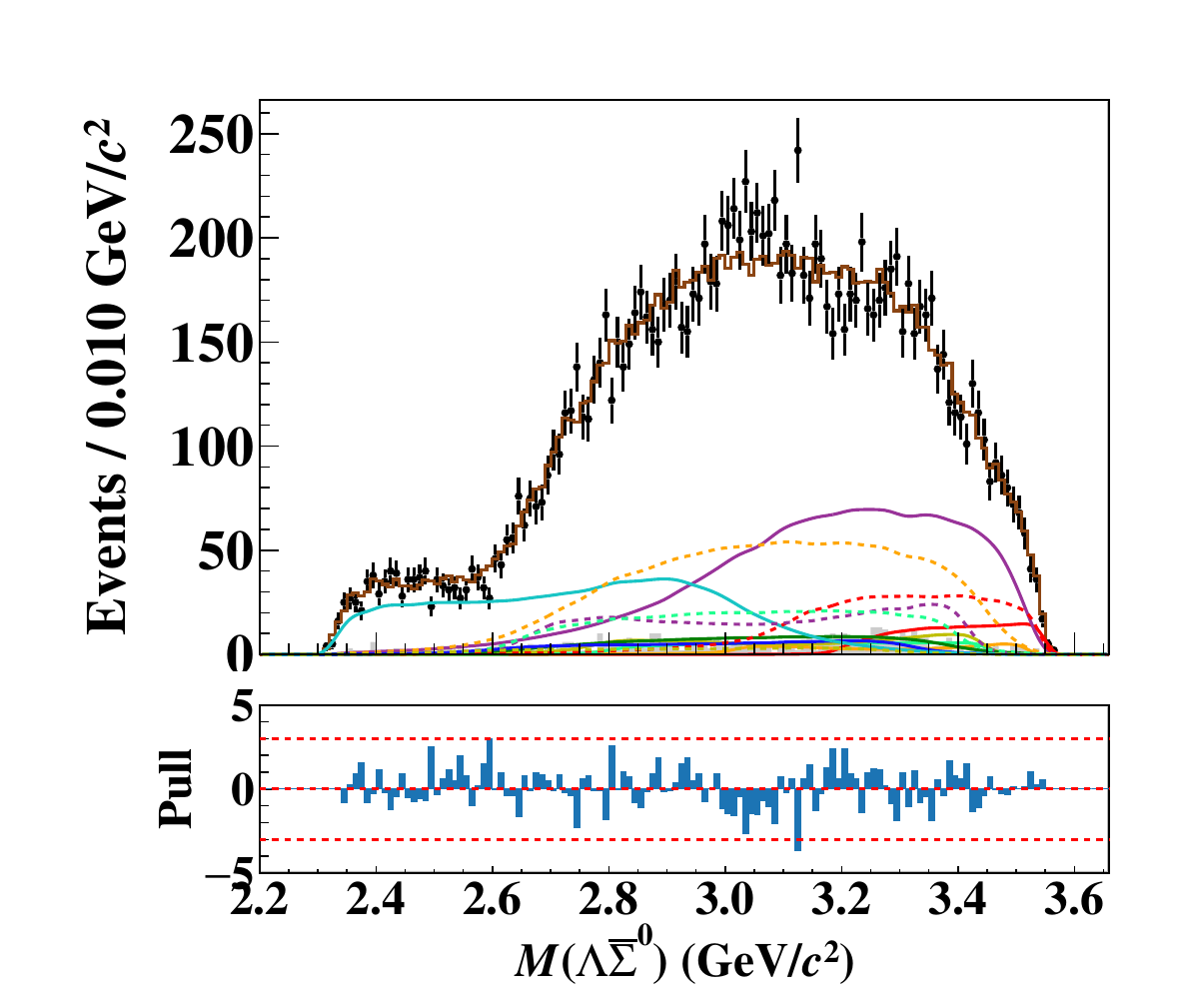}
    \includegraphics[scale=0.3]{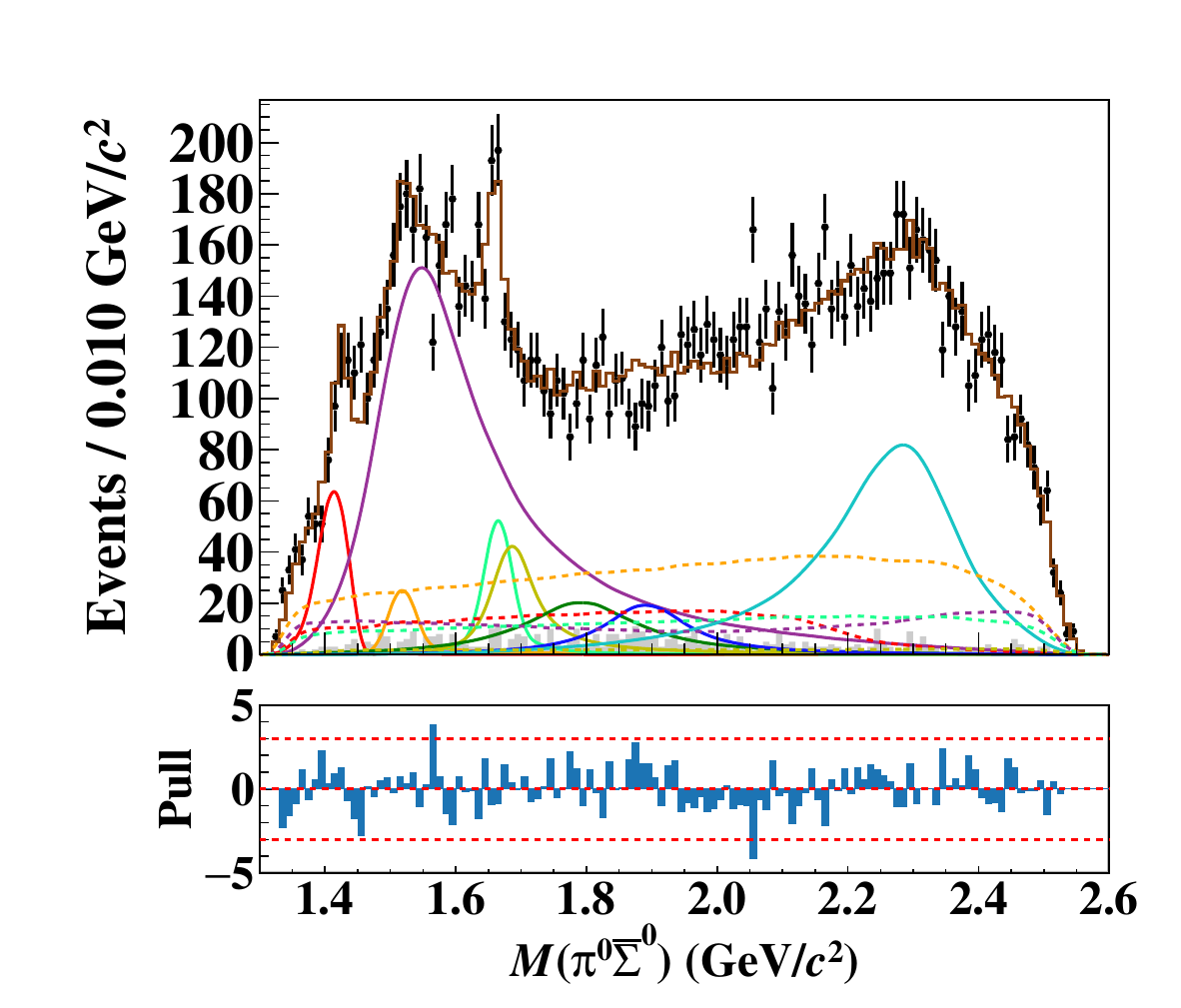}
    \includegraphics[scale=0.3]{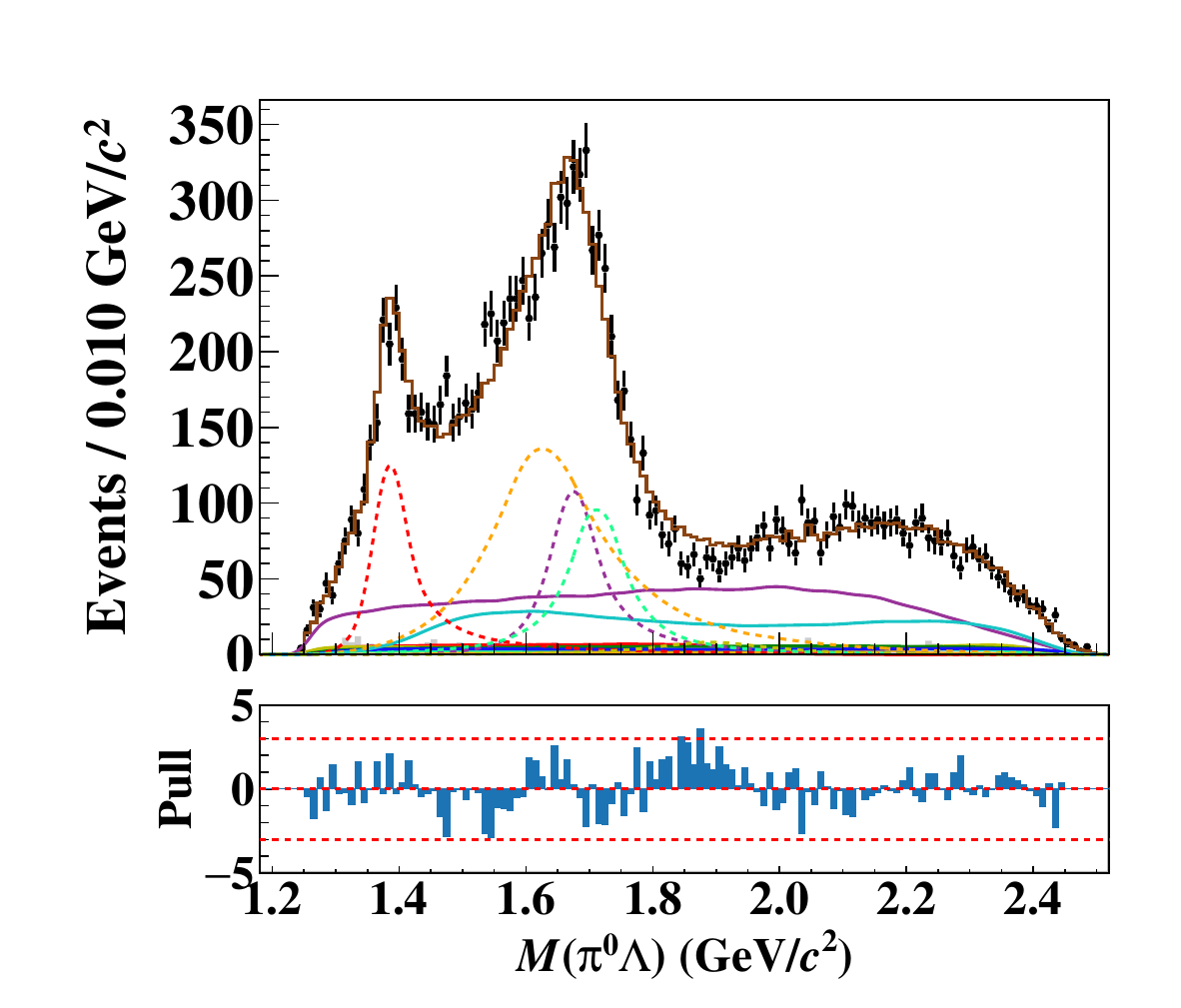}
    \includegraphics[scale=0.3]{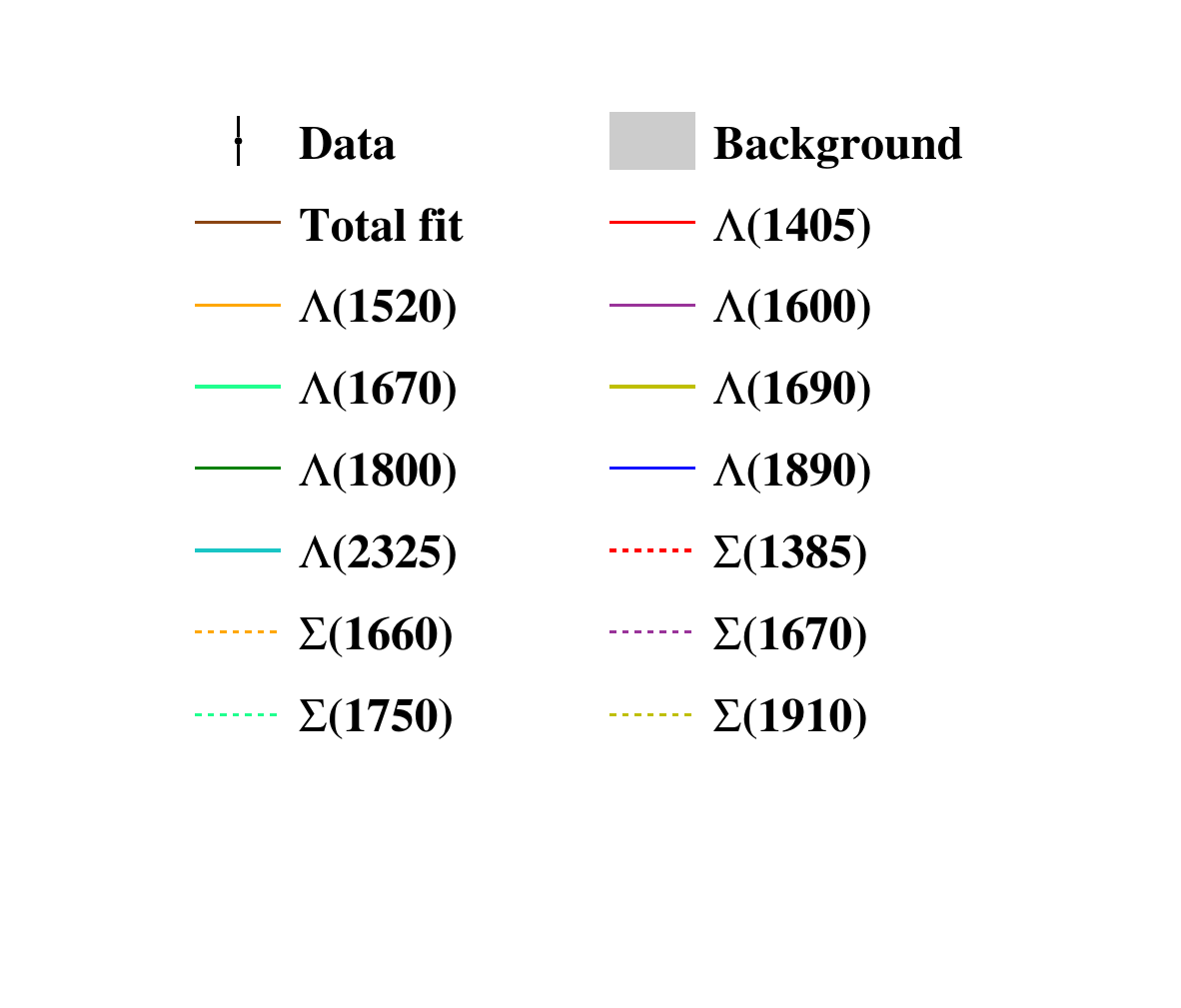}
    \caption{PWA fit curves for $\psi(3686) \to \Lambda \bar{\Sigma}^0 \pi^0$~\cite{BESIII:2024jgy}.}
    \label{fig:pwaLSP}
\end{figure}
\begin{table}[!htp]
    \centering
\caption{Summary of detected $\Lambda^*$ resonance by BESIII collaboration. ``$-$'' mark determines this result is not examined in the analysis, ``$\cdots$'' mark determines this result is evaluated but could not perform the optimal solution.}
\scalebox{0.8}{
    \begin{tabular}{cccccc}
    \hline \hline
    Resonance       &$M$ (MeV/$c^2$)            &$\Gamma$ (MeV)             &$J^P$      &Reference                          &Significance ($\sigma$) \\
    \hline
    $\Lambda(1670)$ &$1672 \pm 5 \pm 6$         &$38 \pm 10 \pm 19$         &$1/2^-$    &\cite{BESIII:2022cxi}         &$>5$ \\
    $\Lambda(2000)$ &$2001 \pm 7$               &$36 \pm 14$                &$-$        &\cite{BESIII:2022fhe}              &$3.0$ \\
    $\Lambda(2325)$ &$2306.5 \pm 6.3 \pm 16.0$  &$227.1 \pm 12.2 \pm 47.8$  &$\cdots$   &\cite{BESIII:2024jgy}              &$>5$ \\
    \hline \hline
    \end{tabular}
}
    \label{SummaryLambda}
\end{table}
\subsection{$\Xi$ and $\Omega$ Baryon}
For $\Xi$ baryon, in 2020, BESIII collaboration had confirmed the $\Xi(1820)$ at the $\Xi$-recoil invariant mass spectra via the cross section study of $e^+ e^- \to \Xi^- \bar{\Xi}^+$ from 4.0 to 4.6 GeV, and determined the mass and width with a Breit-Wigner function; In 2024, both $\Xi(1690)$ and $\Xi(1820)$ are determined in $\psi(3686) \to \Lambda K^- \bar{\Xi}^+$~\cite{BESIII:2023mlv} via PWA method, but the width of $\Xi(1820)$ is much wider than the previous published~\cite{BESIII:2019cuv}. For $\Omega$ baryon, in 2025, BESIII collaboration has reported two $\Omega^*$ resonances~\cite{BESIII:2024eqk} from the $\Omega$-recoil invariant mass spectrum by using the $XYZ$ data sample from 4.13 to 4.70 GeV, one is called $\Omega(2012)$ and has been determined by Belle collaboration in 2018~\cite{Belle:2018mqs}, the other one is firstly determined and to be named as $\Omega(2109)$ with the significance of $4.1\sigma$. The mass and width of $\Omega(2109)$ are also evaluated by the Breit-Wigner function.
\begin{figure}[!htp]
    \centering
    \includegraphics[scale=0.4]{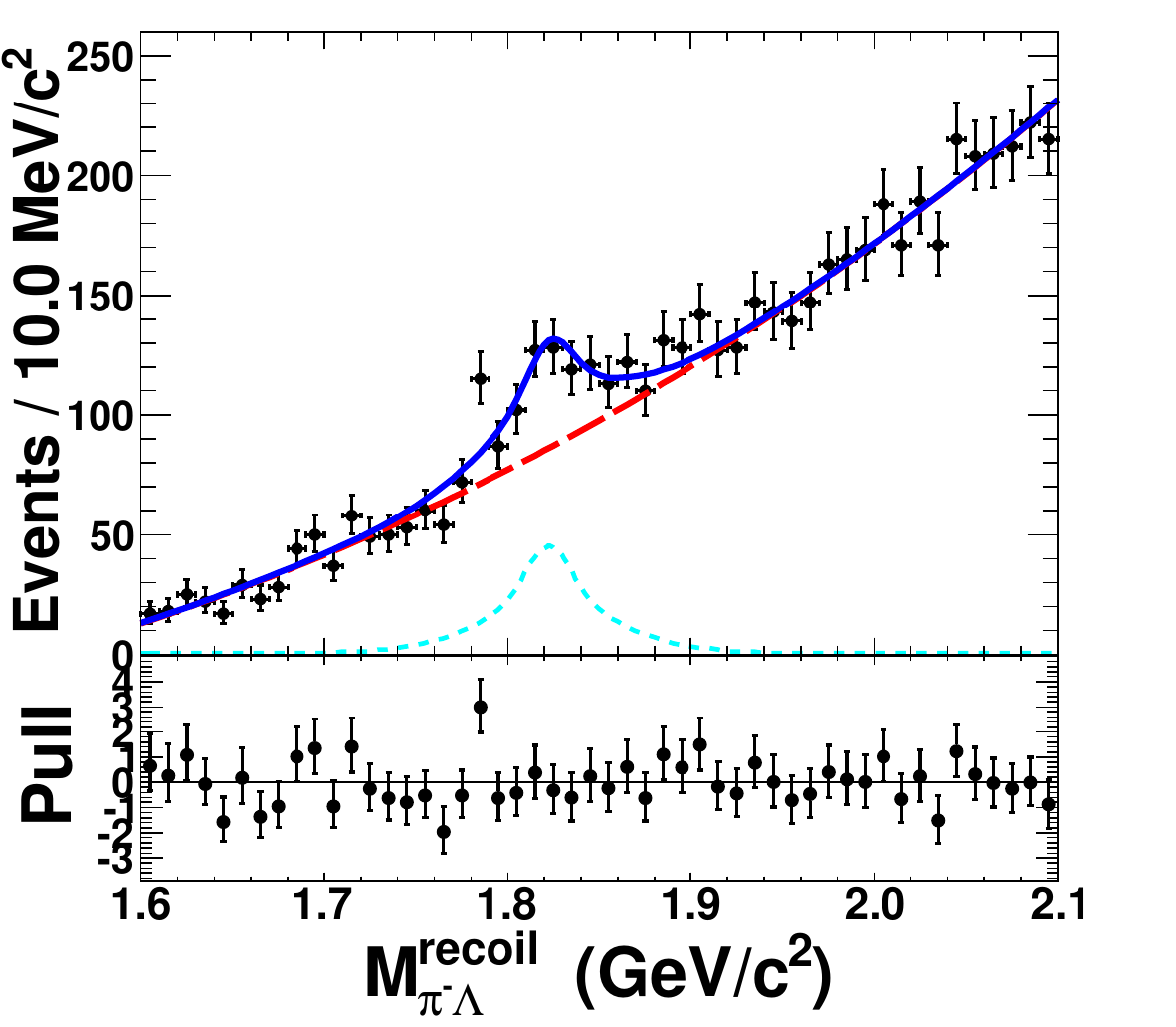}
    \caption{Fit to the recoil mass of $\pi^- \Lambda$ to determine $\Xi(1820)$ by using the combined data, black dots with error bar denote the data, cyan dashed line denotes the signal shape of $\Xi(1820)$ by a Breit-Wigner function convoluted with a Gaussian resolution function, red dashed line denotes the background shape, which is performed by second order Chebyshev polynomial function~\cite{BESIII:2019cuv}.}
    \label{fig:Xi1820Recoil}
\end{figure}
\begin{figure}[!htp]
    \centering
    \includegraphics[scale=0.7]{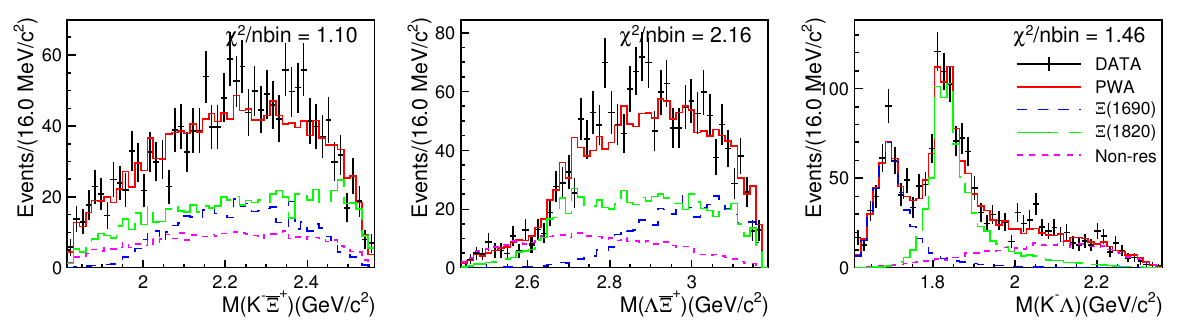}
    \caption{PWA fit curves for $\psi(3686) \to \Lambda K^- \bar{\Xi}^+$~\cite{BESIII:2023mlv}.}
    \label{fig:XiPWA}
\end{figure}
\begin{figure}[!htp]
    \centering
    \includegraphics[scale=0.4]{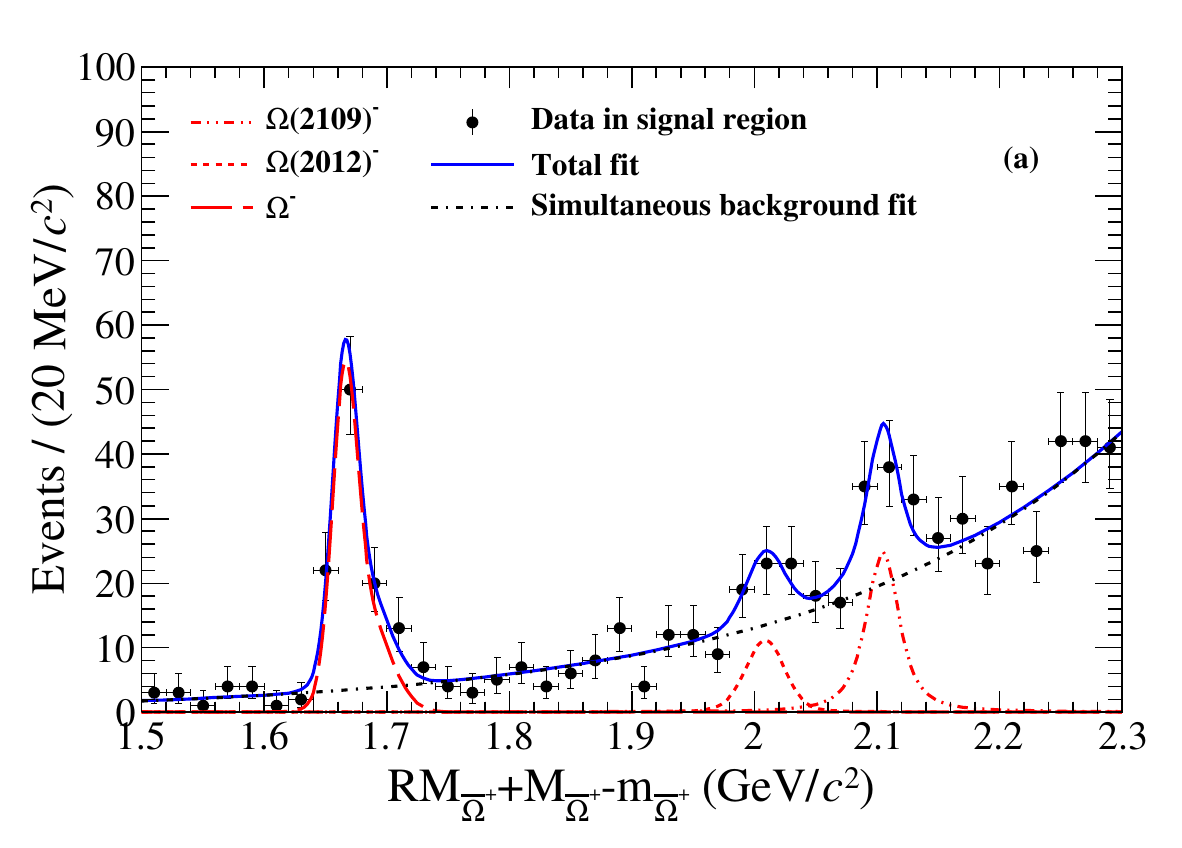}
    \caption{Fit to the recoil mass of $K^- \Lambda$ to determine $\Omega$, $\Omega(2012)$ and $\Omega(2109)$ by using the combined data~\cite{BESIII:2024eqk}.}
    \label{fig:Omega}
\end{figure}
\begin{table}[!htp]
    \centering
\caption{Summary of detected $\Xi^*$ and $\Omega^*$ resonances by BESIII collaboration. ``$-$'' mark determines this result is not examined in the analysis.}
\scalebox{0.8}{
    \begin{tabular}{cccccc}
    \hline \hline
    Resonance       &$M$ (MeV/$c^2$)            &$\Gamma$ (MeV)             &$J^P$      &Reference                          &Significance ($\sigma$) \\
    \hline
    $\Xi(1820)$     &$1825.5 \pm 4.7 \pm 4.7$   &$17.0 \pm 15.0 \pm 7.9$    &$-$        &\cite{BESIII:2019cuv}              &$>6.2$ \\
    $\Xi(1690)$     &$1685^{+3}_{-2} \pm 12$    &$81^{+10}_{-9} \pm 20$     &$1/2^-$    &\cite{BESIII:2023mlv}              &$>>10$ \\
    $\Xi(1820)$     &$1821^{+2}_{-3} \pm 3$     &$73^{+6}_{-5} \pm 9$       &$3/2^-$    &\cite{BESIII:2023mlv}              &$>>10$ \\
    $\Omega(2109)$  &$2108.5 \pm 5.2 \pm 0.9$   &$18.3 \pm 16.4 \pm 5.7$    &$-$        &\cite{BESIII:2024eqk}              &$4.1$ \\
    \hline \hline
    \end{tabular}
}
    \label{SummaryXO}
\end{table}
\section{Summary}
In summary, owing to the large number of $J/\psi$, $\psi(3686)$ and $XYZ$ data samples, as well as the appropriate energy range for the generation of light-excited baryon states, the BESIII collaboration has enriched the baryon spectroscopy study with high precision and additionally discovered several new states; such a result has been accepted in PDG~\cite{ParticleDataGroup:2024cfk} with world-leading precision. Furthermore, baryon study on BESIII is based on charmonia decay, which is a good platform for spectroscopy analysis and violation test of $C\!P$, with powerful PWA tools. These experimental results could help us to deepen our understanding of QCD behavior. Recently, the BESIII Collaboration has decided to update the hardware and continue data collection in the following years, and several interesting analyses are now ongoing as well; expectably more fascinating works will be published in the future.

\section*{Acknowledgement}
This work is supported in part by
the Fundamental Research Funds for the Central Universities under Contracts Nos. lzujbky-2025-ytA05, lzujbky-2025-it06, lzujbky-2024-jdzx06;
National Natural Science Foundation of China under Contracts No. 12247101;
the Natural Science Foundation of Gansu Province under Contracts Nos. 22JR5RA389, 25JRRA799;
the 111 Project under Grant No. B20063.

\end{document}